\documentclass[twocolumn,showpacs,preprintnumbers,superscriptaddress,amsmath,amssymb,floatfix,aps]{revtex4}

\usepackage[dvips]{graphicx}
\usepackage{dcolumn}
\usepackage{bm}
\begin{document}

\title{Influence of water adsorbed on gold on van der Waals/Casimir forces}

\author{G. Palasantzas}
\affiliation{Materials innovation institute M2i and Zernike
Institute for Advanced Materials, University of Groningen,
Nijenborgh 4, 9747 AG Groningen, The Netherlands}

\author{V. B. Svetovoy}
\affiliation{MESA+ Institute for Nanotechnology, University of
Twente, PO 217, 7500 AE Enschede, The Netherlands}

\author{P. J. van Zwol}
\affiliation{Materials innovation institute M2i and Zernike
Institute for Advanced Materials, University of Groningen,
Nijenborgh 4, 9747 AG Groningen, The Netherlands}

\date{\today}

\begin{abstract}
In this paper we investigate the influence of ultra thin water layer
($\sim 1-1.5\; nm$) on the van der Waals/Casimir force between gold
surfaces. Adsorbed water is inevitably present on gold surfaces at
ambient conditions as jump-up-to contact during adhesion experiments
demonstrate. Calculations based on the Lifshitz theory give very
good agreement with the experiment in absence of any water layer for
surface separations $d\gtrsim 10\; nm$. However, a layer of
thickness $h\lesssim 1.5\; nm$ is allowed by the error margin in
force measurements. At shorter separations, $d\lesssim 10\; nm$, the
water layer can have a strong influence as calculations show for
flat surfaces. Nonetheless, in reality the influence of surface
roughness must also be considered, and it can overshadow any water
layer influence at separations comparable to the total sphere-plate
rms roughness $w_{shp}+w$.
\end{abstract}

\maketitle

\section{Introduction}\label{Sec1}
When material objects such as electrodes in
micro/nanoelectromechanical system are separated by distances of
$100\;nm$ or less forces of quantum origin become operative
\cite{Ser97,Cle03,Cha01}. These are the van der Waals (vdW) and
Casimir forces originating from the same physical basis, but having
different names due to historical reasons. The van der Waals force
is the short-distances asymptotic of this general force, for which
one can neglect the retardation of electromagnetic fields, but the
Casimir force is realized at larger distances where the retardation
is important. The common origin of the forces is nicely explained by
the Lifshitz theory \cite{Lif56}, which is able to describe the
transition between the two regimes. It predicts transition between
the Casimir and van der Waals forces at separations $d\sim
\lambda_p/10$, where $\lambda_p=2\pi c/\omega_p$ is the plasma wave
length and $\omega_p$ is the plasma frequency \cite{Gen04,Lam00}.
Keeping in mind the common origin of the forces, in what follows we
will also use a generalized name {\it dispersive forces}.

At separations below $100\;nm$ the Casimir force is very strong and
becomes comparable to electrostatic forces corresponding to voltages
in the range $0.1-1\;V$ \cite{Ser97,Cle03,Cha01}, while for
separations below $10\;nm$ the van der Waals force dominates any
attraction \cite{Tab69,Ede00,Zwo08a}. These properties make the
dispersive force an important player in nanosciencies. Moreover,
from the fundamental point of view, measurements of the forces from
nano to microscales have attracted strong interest in a search of
hypothetical fields beyond the standard model \cite{Ono06}.

In the experiments the dispersive forces are usually measured
between a sphere and a plate. In the vdW regime the force depends on
distance $d$ as $A_H R/6d^2$, where $R$ is the radius of the sphere.
Fits of the experimental data yielded an effective Hamaker constant
$A_H\approx (7-25)\times 10^{-20}\;J$ for gold-water-gold systems
\cite{Ton91}. On the other hand, for $Au$-air-$Au$ surfaces studies
by Tonckt et al. \cite{Ton91} using the surface force apparatus in
the plane-sphere geometry with millimeter size spheres yielded an
effective Hamaker constant of $A_H\approx 28\times 10^{-20}\;J$ for
separations $d>8.5\;nm$. Similar values $A_H\approx 29\times
10^{-20}\;J$ \cite{Pal08} were obtained from AFM force measurements
at separations between $12\;nm$ - $17\;nm$. We are using here the
term "effective Hamaker constant" because in this distance range the
van der Waals asymptotic regime is not fully reached. Fitting the
curve force vs distance following from the Lifshitz theory  in the
range $d=1-5\;nm$ one would find the Hamaker constant to be
$A_H\approx 40\times 10^{-20}\;J$ as it is expected between metal
surfaces \cite{But05}. However, in this distance range the
experimental determination of the constant for $Au$-air-$Au$ is
problematic due to the strong influence of surface roughness and the
strong jump-up-to-contact by formation of capillary bridges due to
adsorbed water \cite{Zwo08d}. It was found that even for the lowest
attainable relative humidity $\sim2\% \pm 1\%$ large capillary
forces are still present.

In this paper we use the term adsorption in sense of physisorption
when the electronic structure of atoms or molecules is barely
perturbed upon adsorption. Formation of capillary bridges takes
place due to water adsorbed on $Au$. We can expect that the surface
of $Au$ is covered with an ultra thin water layer, which is present
on almost all surfaces exposed to air. Experiment \cite{Zwo08d}
suggest that the thickness of this layer is in the nanometer range.
The natural questions one could ask is how thick the water layer is,
and what is the influence of this layers on the dispersive force? At
short separations, $d\lesssim 20\;nm$, these questions become of
crucial importance because they place doubts on our understanding of
the dispersive forces when experiments under ambient conditions are
compared with predictions of the Lifshitz theory. In this paper we
are performing the first steps to answer these questions comparing
experimental data at short separations with the theoretical
predictions for $Au$ covered with a thin layer of water. The problem
is rather nontrivial since  surface wettability can be influenced by
other adsorbates (e.g., hydrocarbons leading to incomplete wetting),
the optical properties of real films must be measured correctly
\cite{Sve08}, as well as the influence of surface roughness has to
be taken into account \cite{Zwo08a,Zwo08b}. The latter lead to
uncertainties in the separation distance of real surfaces, and its
contribution to the force has to be carefully scrutinized
\cite{Zwo08b,Zwo08c}.

The paper is organized as follows. In Sec. \ref{Sec2} we provide
information on surface roughnesses and force measurements in the AFM
experiment, and discuss the distance upon the contact deduced from
the measured roughness. In Sec. \ref{Sec3.1} the main definitions of
the Lifshitz theory are given, and the roughness correction is
related with the measured roughness profile. The dielectric function
of water at the imaginary frequencies is described in Sec.
\ref{Sec3.2}. In Sec. \ref{Sec3.3} we discuss optical data for gold
in relation with the adsorbed water, and in Sec. \ref{Sec3.4} the
results of the dispersive force calculations are given for nonzero
adsorbed water layer. Our conclusions are presented in Sec.
\ref{Sec4}.

\section{Experimental}\label{Sec2}

\begin{figure}[ptb]
\begin{center}
\includegraphics[width=0.45\textwidth]{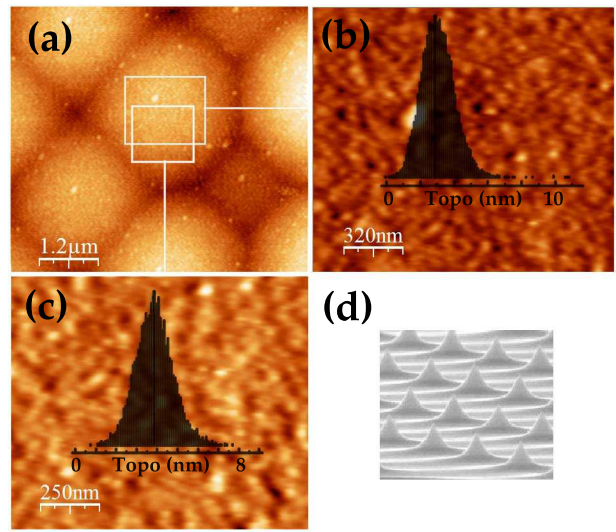}
\vspace{-0.1cm} \caption{(a) Reverse AFM scan of a sphere that has
been used for measurement. (b) Surface scan and and height
distribution with large spots. (c) Surface scan and and height
distribution without large spots. The full width of the roughness
distribution of the sphere without the spots in (c) is about $6\;
nm$. (d) Grid used for the inverse AFM scans of the contact area.}
\label{fig1} \vspace{-0.5cm}
\end{center}
\end{figure}

\begin{figure}[ptb]
\begin{center}
\includegraphics[width=0.45\textwidth]{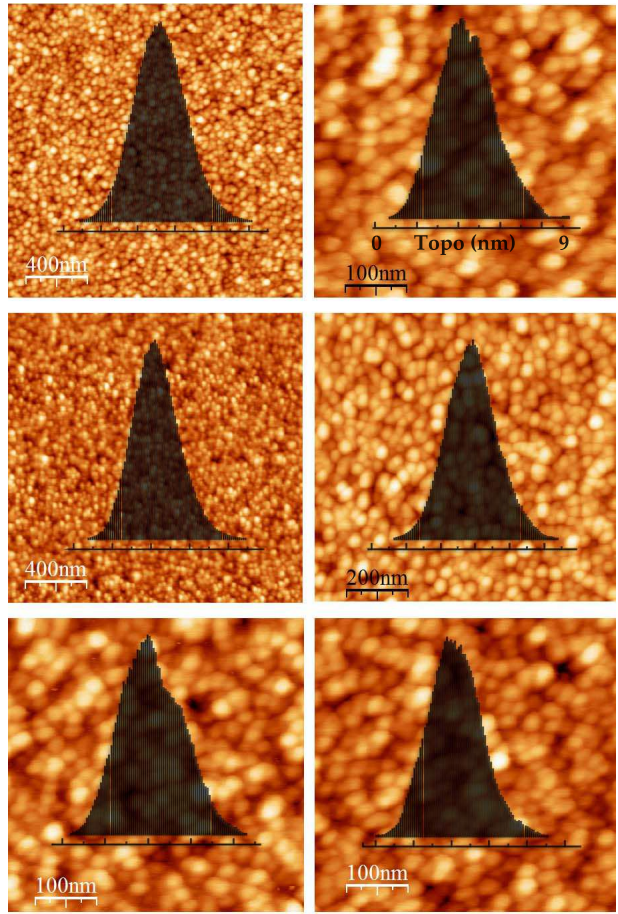}
\vspace{-0.1cm} \caption{Scan areas and height distributions of the
$Au$ films deposited on $Si$ substrate used for the force
measurements.} \label{fig2} \vspace{-0.5cm}
\end{center}
\end{figure}

The dispersive force is measured using the PicoForce AFM
\cite{veeco}, between a sphere with a diameter of $100\;\mu m$
attached on a $240\;\mu m$ long cantilever of stiffness $k=4\; N/m$
(as given by the manufacture), and an $Au$ coated silicon plate.
Both sphere and plate are coated with $100\; nm$ of $Au$, and
afterwards their root-mean-square (rms) roughness was measured by
AFM (see Figs. \ref{fig1} and \ref{fig2}). Analysis of the area on
the sphere where the contact with the substrate occurs was performed
by inverse imaging (Fig. \ref{fig1}) \cite{Pal08,Zwo08c}. The
histograms in Fig. \ref{fig1} and \ref{fig2} show the number of
pixels corresponding to a given height.

Notably the sphere roughness of $1.8\pm 0.2\; nm$ rms \cite{Zwo08b}
is an average over a large area where relatively high spots are
observed (see Fig. \ref{fig1}a,b), which increase the rms roughness
value. For plates surfaces (Fig. \ref{fig2}) these spots were absent
leading to the rms roughness $1.3\pm 0.2\; nm$ \cite{Zwo08b}.
Although the high spots within the contact area may increase the
contact separation between bodies, $d_0$, they deform fast when
pushing surfaces into contact to determine the deflection
sensitivity. This is demonstrated by inverse imaging (see
\cite{Zwo08d} for detailed explanations) and mechanics calculations
also confirm it \cite{Stress}. Locally without the spots or with the
deformed spots the roughness obtained by inverse imaging was
$0.8-1.2\; nm$ rms or $w_{sph}=1\pm 0.2\; nm$ within the contact
area of size  $1\; \mu m$ (Fig. \ref{fig1}c).

Stiff cantilevers used in this experiment do not allow low voltage
electrostatic calibration ($<0.5\; V$) at small separations. To
obtain the contact separation due to roughness $d_0$ we used a
different procedure. One can define the distance between rough
bodies as the distance between zero roughness levels. From the
histograms in Figs. \ref{fig1}, \ref{fig2} it is clear that the zero
roughness level corresponds to the maximum of the distribution.
Without the spots mentioned above these distributions are
approximately symmetric. Two rough surfaces in contact are separated
by the distance $d_0$, which is half of the distance between lowest
and highest points of the rough profile (full width of the
histogram) for the sphere plus the same for the plate. In this way
we found $d_0=7.5\pm 1\; nm$ from multiple AFM scans at different
locations.

It has to be stressed that this direct way (from the definition) to
determine $d_0$ is in good correspondence with what we would expect
from the electrostatic calibration. In Refs. \cite{Zwo08a,Zwo08b}
the relation $d_0\approx (3.7\pm 0.3)\times (w+w_{sph})$ was found
from the electrostatic calibration. In our case the total rms
roughness (sphere and plate) is $w+w_{sph}=1.3+1=2.3\; nm$. Using
this relation we can find $d_0$ within one standard deviation from
that found above from the first principle definition.

\begin{figure}[ptb]
\begin{center}
\includegraphics[width=0.45\textwidth]{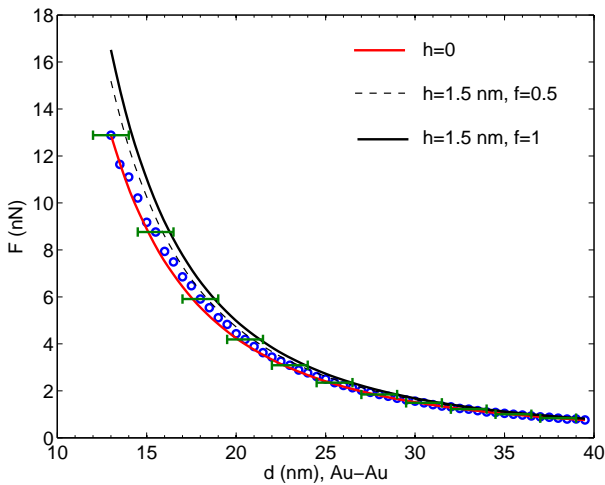}
\vspace{-0.1cm} \caption{Experimental data for the force vs distance
(circles) and theoretical prediction without water layer (red
curve). Errors in the absolute separation are shown for some points
by the bars. The continuous black curve is the prediction for $1.5\;
nm$ continuous water layer. The dashed black curve corresponds to
the same water layer with 50\% of voids.} \label{fig3}
\vspace{-0.5cm}
\end{center}
\end{figure}

The calibration of the deflection sensitivity, cantilever stiffness
$k$, and contact potential $V_0$ was done in the same way as in
previous work \cite{Zwo08a}. Electrostatic fitting in the range of
distances $1-4\;\mu m$ and voltage interval $\pm (3-4.5)\;V$ yielded
the cantilever stiffness $k=8.55\pm 0.38\;N/m$ (with the sphere
attached) and contact potential $V_0=10\pm 10\; mV$. After
calibration, the dispersive force was measured and averaged (using
40 force curves at 20 different locations yielding an average of 800
curves). The result is shown in Fig. \ref{fig3} together with the
theoretical curves (see Sec. \ref{Sec3}) without (red) or with
(black) adsorbed water layer. Each experimental point (circles) is
defined with rather large uncertainty in distance, which are shown
by the horizontal bars for some points. Indeed, the main uncertainty
$\Delta F$ in the dispersive force at short distances comes from the
uncertainty $\Delta d =1\; nm$ in the separation upon contact
$d_0=7.5\pm 1\; nm$. The error in the force due to error in $d_0$
can be estimated as $(\Delta F/F) \approx c(\Delta d_0/d_0) $, where
$c\simeq 2.5$ in the investigated range of distances. Other sources
of errors include the error of the cantilever spring constant
$\Delta k/k\approx 4\%$ and the error in the radius of the sphere
$\Delta R/R=2 \%$. These errors propagated to the force are
negligible in comparison with that arising from $\Delta d_0/d_0$.
Finally, at separations $d>100\; nm$ the force is rather weak, and
the error is dominated by thermomechanical noise as it was explained
in former studies \cite{Zwo08b}.

Evaluation of the force based on the Lifshitz theory requires the
use of the optical properties of the interacting materials as input
data. The optical properties of gold films were measured in air with
ellipsometry in the wavelength range $137\;nm-33\;\mu m$
\cite{Sve08}. Outside of this interval at low frequencies ($\lambda
>33\;\mu m$) the data were extrapolated according to the Drude model
with the plasma frequency $\omega_p=7.84\pm 0.07\; eV$ and the
relaxation frequency $\gamma=49.0\pm 2.1\; meV$. The Drude
parameters were determined from the measured part of the dielectric
function \cite{Sve08}. At high frequencies ($\lambda <137\; nm$) the
method of extrapolation did not play any role, and the data were
taken from the handbook \cite{HB}.

\section{Theory}\label{Sec3}

\subsection{Gold-gold interaction in the Lifshitz theory}\label{Sec3.1}

First we are going to calculate the interaction energy between two
plates $E_{pp}^{flat}$. To be more precise it must be the free
energy if we consider the interaction at finite temperature $T$.
Here, however, we will neglect the thermal effect because our
intention is to calculate the force in the short-distance range $d<
100\; nm$. It is known that for these separations and at room
temperature the thermal correction is small \cite{Mil04}. Physically
it means that the zero-point quantum fluctuations of the
electromagnetic field give the main contribution to the interaction,
while thermally excited fields can be neglected. In this case
instead of summation over the discrete set of Matsubara frequencies,
one can integrate over a continuous imaginary frequency
\cite{Lif56}. The resulting interaction energy corresponding to zero
temperature can be presented in the following form:
\begin{equation}\label{Lif}
    E_{pp}^{flat}(d)=\frac{\hbar}{2\pi}\sum\limits_{\mu}\int
    \limits_0^{\infty}d\zeta\int\frac{d^2q}{(2\pi)^2}
    \ln\left(1-R_{\mu}e^{-2dk_0} \right).
\end{equation}
Here the index $\mu=s,p$ is running two possible polarization states
of the electromagnetic field, and $R_{\mu}$ is the product of the
Fresnel reflection coefficients for plates 1 and 2:
$R_{\mu}=r_{1\mu}r_{2\mu}$. The integration variables have the
physical meaning of the imaginary frequency $\zeta$, and the wave
vector $q=|{\rm{ \bf q}}|$ along the plates.

Here it will be assumed that both interacting surfaces are the same:
$r_{1\mu}=r_{2\mu}=r_{\mu}$. For what follows it will be convenient
to use index {\it 2} for quantities related to gold. If no
additional layer on the gold surface exists, then the Fresnel
coefficients can be written as
\begin{equation}\label{r_def}
    r_s=\frac{k_0-k_2}{k_0+k_2},\ \ \
    r_p=\frac{\varepsilon_2 k_0-k_2}{\varepsilon_2 k_0+k_2},
\end{equation}
where $k_0$ and $k_2$ are defined as the normal components of the
wave vector in vacuum and in $Au$, respectively. These components
are
\begin{equation}\label{k_def}
    k_0=\sqrt{\zeta^2/c^2+q^2},\ \ \
    k_2=\sqrt{\varepsilon_2\zeta^2/c^2+q^2}.
\end{equation}

In Eqs. (\ref{Lif})-(\ref{k_def}) the dielectric function of gold
$\varepsilon_2$ has to be understood as the function at imaginary
frequencies: $\varepsilon_2 = \varepsilon_2(i\zeta)$. This function
cannot be directly measured, but it can be expressed with the
Kramers-Kronig relation via the observable dielectric function
$\varepsilon_2(\omega)$ at real frequencies $\omega$
\begin{equation}\label{K-K}
    \varepsilon_2(i\zeta)=1+\frac{2}{\pi}\int\limits_0^{\infty}d\omega
    \frac{\omega\varepsilon_2^{\prime\prime}(\omega)}{\omega^2+\zeta^2}.
\end{equation}
Note that only the imaginary part of the dielectric function
$\varepsilon_2^{\prime\prime}(\omega)$ contributes to
$\varepsilon_2(i\zeta)$. The fact that the dispersive force depends
on $\varepsilon_2(i\zeta)$ makes the material dependence of the
force sometimes confusing. For example, thin metallic film
transparent for visible light gives very significant contribution to
the force \cite{Sve00a}. On the other hand, a hydrogen-switchable
mirror that changes from reflection to transmission in visible light
does not give a measurable effect \cite{Ian04}.

For practical evaluation of the interaction energy it is convenient
to change variables in (\ref{Lif}). Namely, instead of $q$ one can
introduce $x=2dk_0$. With this variable the integral will run from
$\xi=\zeta/\omega_c$ to $\infty$, where
\begin{equation}\label{omc_def}
    \omega_c=\frac{c}{2d}
\end{equation}
is the characteristic imaginary frequency (it is not a
characteristic frequency in real domain). Because of exponent in the
integrand the integral over $x$ converges fast. However the integral
over $\zeta$ running from 0 to $\infty$ is not convenient for
numerical evaluation. This problem can be solved by substituting
$\zeta=xt\omega_c$. Now $t$ will run from 0 to 1, and in terms of
$x$ and $t$ numerical calculation of the integral in (\ref{Lif})
becomes convenient:
\begin{equation}\label{E_conv}
    E_{pp}^{flat}(d)=\frac{\hbar c}{32\pi^2d^3}\sum\limits_{\mu}
    \int\limits_0^1
    dt\int\limits_0^{\infty}dxx^2\ln\left(1-R_{\mu}e^{-x}\right).
\end{equation}

The roughness correction can be calculated as follows. The force
between a sphere and a plate $F_{sp}$ is related with the
interaction energy per unit area between two plates $E_{pp}$ by the
relation:
\begin{equation}\label{F-E}
    F_{sp}(d)=2\pi R E_{pp}(d),
\end{equation}
where $d$ is the minimal distance between bodies and $R$ is the
sphere radius. Relation (\ref{F-E}) holds true in the limit $R\gg d$
that is the case for our experiment. Roughness gives contribution to
the energy $E_{pp}$, which can be presented as
\begin{equation}\label{Efull}
    E_{pp}=E_{pp}^{flat}+\delta E_{pp}^{rough},
\end{equation}
where the first term corresponds to the interaction energy between
plates.

The second term in Eq. (\ref{Efull}) is responsible for the
roughness correction. It can be presented in the form \cite{Mia05}
\begin{equation}\label{corr}
   \delta E_{pp}^{rough}=\int\frac{d^2k}{(2\pi)^2}G(k,d)\sigma(k).
\end{equation}
Here $\sigma(k)$ is the roughness spectrum and $G(k,d)$ is the
response function derived in \cite{Mia05} both are functions of the
wave number $k$. For self-affine roughness $\sigma(k)$ scales as
\cite{Kri95} $\sigma (k)\propto k^{-2-2H}$ for $k\xi\gg 1$ and
$\sigma (k)\propto const$ for $k\xi\ll 1$. The parameters $\xi$ and
$H$ are the correlation length and roughness exponent, respectively.
For the roughness calculations we used the roughness model in
Fourier space \cite{Pal93}:
\begin{equation}\label{r_model}
    \sigma(k)=\frac{AHw^2\xi^2}{(1+k^2\xi^2)^2},
    \ \ \ A=\frac{2}{\left[1-(1+k_c^2\xi^2)^{-H}\right]},
\end{equation}
where $w$ is the rms roughness, and $k_c\sim 1\;nm^{-1}$ is a lower
roughness cutoff.

For actual calculations we took for the roughness parameters in
$\sigma (k)$ the values: $w_{sph}=1.0\; nm$, $w=1.3\; nm$, lateral
correlation lengths $\xi=20\; nm$, and roughness exponent $H=0.9$
\cite{Zwo08b}. At separations $d>10\; nm$ roughness can still play
significant role by increasing the force. It has to be noted that
relation (\ref{corr}) derived within the scattering theory
\cite{Mia05} is applicable for $d\gg w+w_{sph}$ and small local
surface slopes. For smaller $d$ a much stronger roughness effect is
expected increasing the force up to five times or more with respect
to that of flat surfaces as former studies indicated \cite{Zwo08a}.

We performed calculations of the force between a sphere of radius
$R=50\; \mu m$, and a plate without any water layers using Eqs.
(\ref{E_conv})-(\ref{r_model}). As the dielectric function of $Au$
film we used the data for sample 3 from Ref. \cite{Sve00a}. The
results are presented in Fig. \ref{fig3} by the red line. One can
see that the force without any water layer agrees reasonably well
with the experimental data. The question is what restriction can be
derived from this agreement on the thickness of the water layer on
the gold surface?

\subsection{Dielectric function of water}\label{Sec3.2}

In order to understand how the water layer will contribute to the
force, we have to know first the dielectric function of water at
imaginary frequencies $\varepsilon_1(i\zeta)$. Water is a well
investigated medium, and there are a few works where the dielectric
function was calculated. Parsegian and Weiss \cite{Per81} fitted
$\varepsilon_1(\omega)$ by a number of Lorentzian oscillators as it
is traditionally used by spectroscopists. Then the function
$\varepsilon_1(i\zeta)$ can be found by analytic continuation. This
method, however, does not always give sufficient precision. For
example, refitting of the same input data used in \cite{Per81} gave
considerably different result \cite{Rot96}. Recently a new
analytical model for the dielectric function was proposed
\cite{Shu07}.

A more reliable approach can be based on the direct use of the
optical data of water. It was realized in Ref. \cite{Dag00} where
$\varepsilon_1(i\zeta)$ was found from available optical data in a
wide range of frequencies. The authors were directed to calculation
of the van der Waals force at rather small separations $d\sim
1\;nm$. The important imaginary frequencies where
$\varepsilon_1(i\zeta)$ has to be known with the best possible
precision are around $\zeta\sim\omega_c=c/2d$. For $d\sim 1\; nm$
important frequencies are $\zeta\sim 100\; eV$. To have
$\varepsilon_1(i\zeta)$ in this frequency range one has to integrate
in the dispersion relation (\ref{K-K}) (but for $\varepsilon_1$) at
$\omega\gtrsim 100\; eV$. For these high frequencies the directly
measured quantity is ${\rm Im}\left(1/\varepsilon(\omega)\right)$.
The real part of $1/\varepsilon(\omega)$ must be restored with the
Kramers-Kronig relation. The complete procedure for calculation of
$\varepsilon(i\zeta)$ is rather complex \cite{Dag00}. In our case,
for $d\gtrsim 10\;nm$ this procedure can be significantly
simplified.

Segelstein \cite{Seg81} compiled the data for the absorption
coefficient of water in very wide range of wavelengths from $10\;
nm$ to $1\; m$. These data can be used instead of
$\varepsilon_1^{\prime\prime}(\omega)$ to calculate
$\varepsilon_1(i\zeta)$. The absorption coefficient $\mu(\omega)$ is
related with the imaginary part of the complex refractive index
$\tilde{n}(\omega)=n(\omega)+ik(\omega)$ by the relation
$\mu(\omega)=2\omega k(\omega)/c$. Between $n(\omega)$ and
$k(\omega)$ exists similar dispersion relation as between
$\varepsilon_1^{\prime}(\omega)$ and
$\varepsilon_1^{\prime\prime}(\omega)$:
\begin{equation}\label{refr}
    \tilde{n}(i\zeta)=1+\frac{2}{\pi}\int\limits_0^{\infty}d\omega
    \frac{\omega k(\omega)}{\omega^2+\zeta^2}.
\end{equation}
If we know $\tilde{n}(i\zeta)$ then $\varepsilon_1(i\zeta)$ can be
expressed as $\varepsilon_1(i\zeta)=\tilde{n}^2(i\zeta)$. This is
true because both functions $\tilde{n}$ and $\varepsilon_1$ are
analytical.

\begin{figure}[ptb]
\begin{center}
\includegraphics[width=0.45\textwidth]{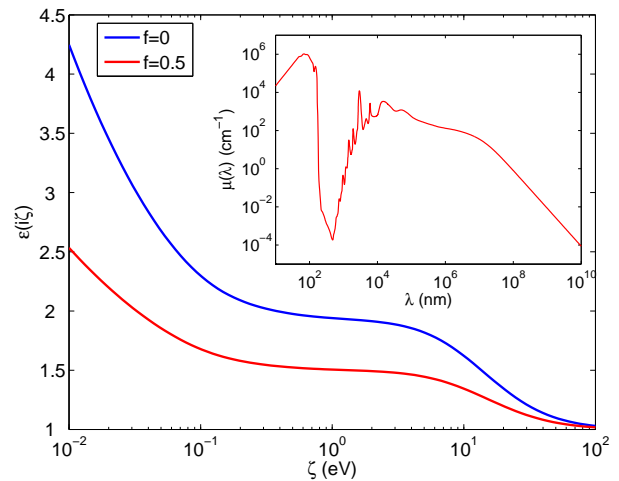}
\vspace{-0.1cm} \caption{The dielectric function of water at
imaginary frequencies. The blue curve is for bulk water (no free
volume, $f=0$). The red curve corresponds to water with 50\% of free
volume, $f=0.5$ (see Sec. \ref{Sec3.4}). The inset shows the
compiled data for absorption coefficient used as input data to
calculate $\varepsilon_1(i\zeta)$.} \label{fig4} \vspace{-0.5cm}
\end{center}
\end{figure}

The resulting dielectric function of water calculated at the
imaginary frequencies is presented in Fig. \ref{fig4}. The inset
shows the compiled data \cite{Seg81} for the absorption coefficient
of water $\mu(\lambda)$. Our result is close to that in Ref.
\cite{Dag00} and deviates from both in \cite{Per81} and
\cite{Rot96}. It is interesting to note that $\varepsilon_1(i\zeta)$
is still far from its static value $\varepsilon_1(0)\approx 80$ even
at $\zeta=0.01\; eV$. The static value is reached only at $\zeta
\sim 10^{-6}\; eV$. This is a specific property of water.

\subsection{Water film and optical data}\label{Sec3.3}

The optical properties of our gold films were measured at ambient
conditions \cite{Sve08}. An ultrathin film of water is already
incorporated in the optical response of these films. It means that
the force calculated with the dielectric function of nominal gold
could already include some effect of water. Then the agreement
between forces measured and predicted theoretically for pure gold
becomes questionable. Therefore, the first question to answer is how
well we may know the dielectric function of gold samples from
measurements in ambient conditions? This question has also an
independent interest for the community dealing with the Casimir
effect. Moreover, one could try to extract information on thickness
of water films from optical measurements. These problems are closely
related and we will analyze them in this subsection.

The ellipsometry can give the "pseudodielectric" function of the
investigated material as
\begin{equation}\label{ps_diel}
    \langle \varepsilon\rangle=\sin^2\vartheta\left[1+\tan^2\vartheta
    \left(\frac{1-\rho}{1+\rho}\right)^2\right],
\end{equation}
where $\vartheta$ is the angle of incidence, and $\rho$ is the
directly measured complex ratio of the reflection coefficients
$\rho=r_p/r_s$. If the investigated material is a pure gold film,
then $\langle \varepsilon\rangle$ as calculated from (\ref{ps_diel})
will correspond to the dielectric function of $Au$ and the
reflection coefficients $r_{s,p}$ will coincide with that in
(\ref{r_def}).

If there is a water layer of thickness $h$ on top of gold film, then
the reflection coefficients entering in (\ref{ps_diel}) will be
different. Let us enumerate the media air-water-gold by increasing
indexes 0-1-2. Then for both polarizations we have
\begin{equation}\label{r_multi}
    r=\frac{r_{01}-r_{21}e^{2ik_1h}}{1-r_{01}r_{21}e^{2ik_1h}},
\end{equation}
where the reflection coefficients $r_{ij}$ on the border between
media $i$ and $j$ are defined as
\begin{equation}\label{rij}
    r_{s,ij}=\frac{k_i-k_j}{k_i+k_j},\ \ \ r_{p,ij}=
    \frac{\varepsilon_jk_i-\varepsilon_ik_j}
    {\varepsilon_jk_i+\varepsilon_ik_j}.
\end{equation}
Since we expect that the thickness of the water layer is small
($\sim 1\; nm$) then the phase factor $k_1h$ will be small
$|k_1h|\ll 1$ at all frequencies covered by the ellipsometers. This
is true because in the investigated frequency range
$|\varepsilon_1(\omega)|\lesssim 2$. For this reason the measured
"pseudodielectric" function $\langle\varepsilon\rangle$ only
slightly deviates form the dielectric function of gold
$\varepsilon_2$:
\begin{equation}\label{de}
    \langle\varepsilon\rangle=\varepsilon_2+
    \delta\varepsilon,\ \ \
    \left|\frac{\delta\varepsilon}{\varepsilon_2}\right|\ll 1.
\end{equation}
Using the perturbation theory in $\delta\varepsilon$ one can find
from (\ref{ps_diel})-(\ref{rij}) the expression for the relative
correction to the dielectric function of $Au$:
\begin{equation}\label{deps}
    \frac{\delta\varepsilon}{\varepsilon_2}=i\frac{4\pi h}{\lambda}
    \sqrt{\langle \varepsilon\rangle-\sin^2\vartheta}\;
    \frac{ \langle \varepsilon \rangle - \varepsilon_1 }
    {\langle \varepsilon \rangle-1}\cdot
    \frac{\varepsilon_1-1}{\varepsilon_1}.
\end{equation}
This equation precisely coincides with that presented by Aspnes in
Ref. \cite{Asp82}.

If the dielectric functions of gold $\varepsilon_2$ and water
$\varepsilon_1$ are known, then using Eqs. (\ref{de}) and
(\ref{deps}) one could find the film thickness $h$. This can be
performed by the best fit of the known $\varepsilon_2$ with the
calculated $\langle \varepsilon\rangle-\delta\varepsilon$. However,
as it was demonstrated \cite{Sve08} the optical properties of opaque
$Au$ films depend on the preparation method on the level, which
cannot be ignored. For this reason $\varepsilon_2$ is essentially
unknown.

The relative correction (\ref{deps}) is not negligible at low and
high frequencies. At low frequencies the ratio $h/\lambda$ is small
but $ \langle \varepsilon \rangle $ is large as it is the case for
all good conducting materials. At the largest investigated
wavelength $\lambda=30\; \mu m$ and $h=1\; nm$, the correction is
about 3\%. However, for $\lambda>20\; \mu m $ the noise in the data
becomes significant, and this correction is below the noise level.
If $h$ is larger, we could in principle determine its value. This is
because at low frequencies $\varepsilon_2$ can be described by the
Drude model, and we can determine $h$ together with the Drude
parameters $\omega_p$ and $\gamma$. This procedure was applied for
all investigated $Au$ films but minimization gave unreasonable
values of the parameters including a negative water layer thickness.
Moreover, the target function used in the minimization is larger
than that in the case $h=0$. This probably means that $h$ is small
enough so that the correction $\delta \varepsilon $ is on the noise
level.

In the high frequency limit the correction is defined by the ratio
$h/\lambda$, which is not very small. This frequency range
corresponds to the interband absorption of gold where the dielectric
function $\varepsilon_2$ cannot be predicted. For this reason we
cannot determine $h$ because we have no reliable value to compare
with $\langle \varepsilon\rangle-\delta\varepsilon$.

We can conclude that the dielectric function measured
ellipsometrically is a good approximation for the dielectric
function of pure gold even if there is a layer of adsorbed water.
Information on the thickness of the water layer $h\sim 1\;nm$ cannot
be extracted from the data since it is below the noise level.

\subsection{The dispersive force for nonzero water layer}\label{Sec3.4}

\begin{figure}[ptb]
\begin{center}
\includegraphics[width=0.45\textwidth]{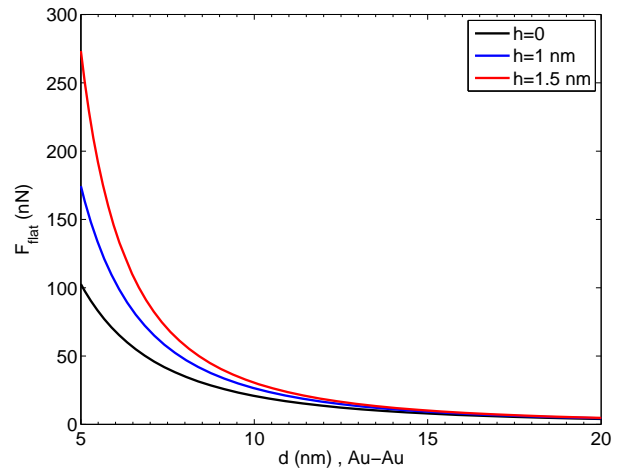}
\vspace{-0.1cm} \caption{The force between a flat sphere ($R=50\;\mu
m$) and a plate in the range of small separations between gold
surfaces. The curves correspond to different thicknesses $h$ of the
water layer.} \label{fig5} \vspace{-0.5cm}
\end{center}
\end{figure}

In the evaluation of the force it will be assumed that water forms a
continuous film on the gold surface. In general it can be not the
case because water can wet metal surface incompletely due to
presence of hydrocarbons and other chemicals on the surface. In the
case of incomplete wetting one can use the approach of effective
dielectric functions \cite{Asp82}, which reduces the problem to a
continuous layer with an effective dielectric function. If
$\varepsilon_1$ is the dielectric function of water in its
homogeneous form, then the effective dielectric function
$\tilde{\varepsilon}_1$ of the material containing a volume fraction
of voids $f$ can be found from the equation
\begin{equation}
\frac{\tilde{\varepsilon}_1 -\varepsilon _{H}}{
\tilde{\varepsilon}_1 +2\varepsilon _{H}}=f\frac{1-\varepsilon _{H}}{%
1+2\varepsilon _{H}}+(1-f)\frac{\varepsilon_1 -\varepsilon _{H}}{%
\varepsilon_1 +2\varepsilon _{H}},  \label{EMA}
\end{equation}
where $\varepsilon _{H}$ is the dielectric function of the ''host''
material. In the Bruggman approximation it is assumed that the host
material coincides with the effective medium, $\varepsilon
_{H}=\tilde{\varepsilon}_1 $, so it treats both void and material
phases on an equal basis. At the imaginary frequencies
$\tilde{\varepsilon}_1$ with 50\% of voids ($f=0.5$) is shown in
Fig. \ref{fig4} as the red line.

We can calculate the force between flat surfaces using the formulas
(\ref{E_conv}) and (\ref{F-E}). Now for the reflection coefficients
we have to use Eqs. (\ref{r_multi}) and (\ref{rij}) taken at
imaginary frequencies.

Figure \ref{fig5} shows the force at small distances between $Au$
surfaces when both of them are covered with a continuous water layer
of thickness $h=0,\;1,\;{\rm and}\;1.5\;nm$. Indeed, the force
versus distance between $Au$ surfaces increases with the water
layer. This is because the external boundaries of the bodies (water
surfaces) are separated by the smaller distance $d-2h$. As one can
see the effect of water becomes very significant at separations $d$
below $10\;nm$, which are not accessible in this study. We presented
only the forces between flat surfaces because at these small
separations there is no a reliable way to estimate the roughness
correction. The method developed in \cite{Mia05} can be applied only
for $d\gg w+w_{sph}$; for the system under investigation this
condition is $d\gg 2.3\;nm$. At distances where the theory is not
applicable it was demonstrated experimentally \cite{Zwo08a} that the
roughness correction increases very significantly, although it was
proven for much rougher surfaces than those in the present study.

The effect of water layer on the dispersive force for experimentally
investigated separations is shown in Fig. \ref{fig3} by continuous
and dashed black curves, where we included also the effect of
roughness. As one can see a continuous layer of a thickness of
$h=1.5\; nm$ is already excluded by the experiment on the level of
one standard deviation. However, if there is a discontinuous layer
of water containing 50\% of air, it is still in agreement with the
experiment. We can conclude that the effect of water layer on the
dispersive force is considerably masked by large error in the
separation upon contact $d_0$. On the other hand, for the distances
$d<10\; nm$ the force measurements are rather limited due to strong
jump-up-to contact by formation of capillary bridges \cite{Zwo08d}.
In addition for these distances not only water layer but also
roughness is a significant factor, which increases the force up to
five times with respect to flat surfaces as our former studies
\cite{Zwo08a} indicated. All this lead to a rather complex situation
at $d<10\; nm$.

\section{Conclusions}\label{Sec4}

In conclusion, we investigated the influence of an ultra thin water
layer on the dispersive force between $Au$ surfaces. Evaluation of
the force in terms of the Lifshitz theory predicts a strong
influence of the water layer on the force, especially for small
separations $d<10\; nm$, where a higher force is obtained because
the effective distance between the $Au$ surfaces decreases.
Furthermore, the theoretical predictions are compared to the
experimental measurements for distances $d\geq 13\; nm$ (limited due
to strong jump-up-to contact by formation of capillary bridges). It
is shown that although the water layer increases the force, it falls
within the error margins of the measured force. The errors are shown
to arise mainly from the experimental uncertainty in determining the
separation upon contact due to nanoscale surface roughness.

Notably at short separations (comparable to the total sphere-plate
rms roughness $w_{sph}+w$) the influence of surface roughness is
also significant. The roughness can also strongly increase the force
that complicates the situation to a significant degree. In any case,
further experimental work is necessary in combination with smoother
surfaces to minimize roughness contributions. Roughness is a rather
strong barrier because the measurements we did already are at the
limits of realistically possible sphere smoothness. The influence of
the water layer will be weaker if it wets incompletely the metal
surfaces. It can happen due to presence of hydrocarbons and other
chemicals leading to a relatively large contact angles $80^{\circ}$
\cite{But05,Zwo08d}.

{\bf Acknowledgements} The research was carried out under project
number MC3.05242 in the framework of the Strategic Research
programme of the Materials innovation institute M2i (the former
Netherlands Institute for Metals Research (NIMR)). Financial support
from the Materials innovation institute M2i is gratefully
acknowledged.



\begin{thebibliography}{99}
\bibitem{Stress} The yield stress for $Au$ is $<100\; Mpa$ (the ultimate
strength of $Au$). A $4\;N/m$ cantilever bends $100\;nm$ after
contact with the surface (to dermine the deflection sensitivity $m$)
exerts a force of $400\; nN$ onto the surface upon contact. The
pressure on a local spot of size say $50\; nm$ is about $160\;Mpa$,
which is above the ultimate stress for $Au$. Therefore, the features
will deform and they do not pose a problem for determing $d_0$.
\end{thebibliography}
\end{document}